\documentclass[12pt]{article}
\usepackage{epsfig,wrapfig}
\newcommand{\beq}{\begin{equation}}
\newcommand{\eeq}{\end{equation}}
\newcommand{\bea}{\begin{eqnarray}}
\newcommand{\eea}{\end{eqnarray}}
\newcommand{\Veff}%
{{\cal V}^{\rm\scriptsize p}_{\rm\scriptsize eff}}
\newcommand{\Fs}{\rm\scriptsize F}
\newcommand{\Go}{\rm\scriptsize G}
\newcommand{\Pa}{\rm\scriptsize P}
\begin{document}

\begin{center}{\Large\bf Surface behaviour of the pairing gap \\
in a slab of nuclear matter }
\end{center}

\vskip 1 cm

\centerline{\large M.~Baldo$^{1}$, M.~Farine$^{2}$,
U.~Lombardo$^{3,4}$, E.~E.~Saperstein$^{5}$, P.~Schuck$^{6}$,}

\vskip 0.1 cm
\centerline {\large and M.~V.~Zverev$^{5}$}

\vskip 1 cm

\begin{center}
{\small\sl $^1$INFN, Sezione di Catania, 57 Corso Italia,
I-95129 Catania, Italy \\
$^2$Ecole Navale, Lanv\'eoc-Poulmic, 29240 Brest-Naval,
France \\
$^3$INFN-LNS, 44 Via S.-Sofia, I-95123 Catania, Italy \\
$^4$ Dipartimento di Fisica, 57 Corso Italia,
I-95129 Catania, Italy \\
$^5$ Kurchatov Institute, 123182, Moscow, Russia \\
$^6$Insitut de Physique Nucl\'eaire, IN2P3-CNRS, \\
Universit\'e Paris-Sud, F-91406 Orsay-C\'edex, France }
\end{center}

\vskip 3.5 cm

\hrule \vskip 0.5 cm

{\bf Abstract} \vskip .5 cm

\begin{small}
The surface behavior of the pairing gap previously
studied for semi-infinite nuclear matter is analyzed in
the slab geometry. The gap-shape function is calculated
in two cases: (a) pairing with the Gogny force
in a hard-wall potential and (b) pairing with the
separable Paris interaction in a Saxon-Woods mean-field
potential. It is shown that the surface features are preserved
in the case of  slab geometry, being almost independent of
the width of the slab. It is also demonstrated that
the surface enhancement is strengthened as the absolute
value of chemical potential $|\mu|$ decreases which simulates
the approach to the nucleon drip line.
\end{small}

\vskip 0.5 cm \hrule

\newpage

\centerline {\large\bf 1. Introduction}
\vskip .3 cm

Recently,
the surface behavior of the pairing gap in the
$^1S_0$-channel in semi-infinite nuclear matter was
investigated independently from two quite
different approaches \cite{BLSZ1}, \cite{FS}.
A rather sophisticated approach was used in \cite{BLSZ1}
which starts from the microscopic gap equation
for semi-infinite nuclear matter with
the separable representation
\cite{Par1,Par2} of the Paris potential
\cite{Paris}.
The effective pairing interaction $\Veff$ adopted in the gap equation
 was previously found within the Bethe-Goldstone formalism for
semi-infinite nuclear matter
without any form of local approximation \cite{BLSZ2}.
All the calculations were made for two values of the chemical potential:
$\mu{=}-16\,$MeV and $\mu{=}-8\,$MeV.
A surface enhancement in the gap $\Delta$ was found,
the effect being more pronounced for $\mu=-8$\,MeV.

In \cite{FS} a more simplified model was used
in which nuclear matter was embedded in a semi-infinite
hard wall potential and the pairing problem was
considered in the BCS approximation with the Gogny force.
Such a simple approach makes it possible to solve
the problem to a great deal analytically and to examine the coordinate
dependence of the pairing gap, pairing tensor and
correlation energy density in a rather transparent way.
A relatively mild surface enhancement of all the quantities under
consideration was found. As to $\Delta$, it is of the same size
as in \cite{BLSZ1}.

In this paper we carried out an analogous analysis for a slab
of nuclear matter within  both approaches
with the hope that a direct comparison of results can help
to clarify the general features of the phenomenon under
consideration. The slab system is much closer to real
atomic nuclei than the semi-infinite one and many results
can be qualitatively related to them.

The structure of the article is as follows. In Section 2 we
extend the model with the Gogny force \cite{FS} to the case
of a hard-wall slab potential. Section 3 contains the extension
of the model of \cite{BLSZ1} with the Paris force to the case
of slab geometry. The results obtained in both
models are discussed in Section 4.

\vskip .5 cm
\centerline {\large\bf 2. Pairing with the Gogny force in the
hard wall slab potential.}
\vskip .3 cm

Let us consider a slab of nuclear matter embedded in a hard wall
potential of thickness $2L$ along the $x$-direction. Let the
origin be at the center of the slab. We start from expanding
the gap operator in $r$-space $\Delta({\bf r},{\bf r}')$ in terms of the
wave functions $\varphi_{\bf k}({\bf r)}$ of
the hard-wall slab potential,
\beq
\varphi_{\bf k}({\bf r})=
{1\over L}\,\theta(L{+}x)\,\theta(L{-}x)\,\sin k_x (x{-}L)\,
e^{i{\bf k}_{\perp}{\bf r}_{\perp}},
\label{phi}
\eeq
where $k=\{k_x,{\bf k_{\perp}}\}$, the quantum number $k_x$ running
over the discrete set of eigenvalues $k_n=\pi n/(2L)$, $n=1, 2, \dots$.
Within the usual BCS approximation the expansion reads
\beq
\Delta({\bf r}_1,{\bf r}_2)=\sum\limits_{\bf k}
\varphi_{\bf k}({\bf r}_1)\,\varphi_{-{\bf k}}({\bf r}_2)\,\Delta({\bf k})\, ,
\label{expan}
\eeq
where the state $|-{\bf k}\bigr>$ is time-reversed with
respect to the state $|{\bf k}\bigr>$.
The gap $\Delta({\bf k})$ obeys the BCS equation
\beq
\Delta({\bf p})=-\sum\limits_{\bf k}
\bigl<{\bf p},-{\bf p}\bigl|V\bigr|-{\bf k},{\bf k}\bigr>
\,\frac{\Delta({\bf k})}{2\sqrt{\xi^2({\bf k})+\Delta^2({\bf k})}} \, ,
\label{BCS}
\eeq
where $\xi({\bf k})=\varepsilon({\bf k})-\mu$ is the
single-particle energy relative to the chemical potential
and
\beq
\bigl<{\bf p},-{\bf p}\bigl|V\bigr|-{\bf k},{\bf k}\bigr>
= \int\!\!\!\int d{\bf r}_1\,d{\bf r}_2\,
\varphi^*_{\bf p}({\bf r}_1)\,\varphi^*_{-{\bf p}}({\bf r}_2)\,
V({\bf r}_1,{\bf r}_2)\,
\varphi_{-{\bf k}}({\bf r}_2)\,\varphi_{\bf k}({\bf r}_1)
\label{matel}
\eeq
are the matrix elements of the pairing interaction
$V({\bf r}_1,{\bf r}_2)$. For the purpose of performing calculations
analytically we used here, as in \cite{FS}, the Gogny force D1
\beq
V({\bf r_1},{\bf r_2})=\sum\limits_{c=1}^2 (W_c-B_c-H_c+M_c)\,
e^{-({\bf r_1}-{\bf r_2})^2/\alpha_c^2}
\label{gogny}
\eeq
with the values of the parameters given in \cite{dechgog}.

Using the explicit form of the eigenfunctions (\ref{phi})
in Eq.~(\ref{matel}) one obtains
$$
\bigl<{\bf p},-{\bf p}\bigl|V\bigr|-{\bf k},{\bf k}\bigr>
={1\over 2L^2}\sum\limits_{c=1}^2 \,(W_c-B_c-H_c+M_c)\,
\pi\,\alpha_c^2\,
e^{-\alpha_c^2(p^2_{\perp}-k^2_{\perp})/4}
\qquad\qquad
$$
$$
\times \int\!\!\!\int dx_1\,dx_2\, \theta(x_1{+}L)\,\theta(L{-}x_1)\,
 \theta(x_2{+}L)\,\theta(L{-}x_2)
\qquad\qquad
$$
\beq
\times \sin\Bigl(k_x(x_1{-}L)\Bigr)\,\sin\Bigl(k_x(x_2{-}L)\Bigr)\,
\sin\Bigl(p_x(x_1{-}L)\Bigr)\,\sin\Bigl(p_x(x_2{-}L)\Bigr)\,
e^{-(x_1-x_2)^2/\alpha_c^2}\, .
\label{matel1}
\eeq

Upon substituting this expression into Eq.~(\ref{BCS}) and,
as the $s$-wave pairing is considered, averaging over
the angle between vectors ${\bf p}$ and ${\bf k}$,
the gap equation can be rewritten as follows
$$
\Delta(p_m,p_{\perp})
={1\over 4L^2}\sum\limits_{c=1}^2\, (W_c-B_c-H_c+M_c)\,
\alpha_c^2\,
e^{-\alpha_c^2 p^2_{\perp}/4}
\qquad\qquad\qquad\qquad
$$
$$
\times
\sum\limits_n\int\limits_0^{\infty} k_{\perp}\, dk_{\perp}
e^{-\alpha_c^2 k^2_{\perp}/4}
I_0\left({\alpha_c p_{\perp} k_{\perp}\over 2} \right)
\,\frac{\Delta(k_n,k_{\perp})}
{\sqrt{\xi^2(k_n,k_{\perp})+\Delta^2(k_n,k_{\perp})}}
$$
$$
\times \int\int dx_1dx_2 \theta(x_1{+}L)\,\theta(L{-}x_1)\,
 \theta(x_2{+}L)\,\theta(L{-}x_2)
$$
\beq
\times \sin\Bigl((k_n(x_1{-}L)\Bigr)\,\sin\Bigl(k_n(x_2{-}L)\Bigr)\,
\sin\Bigl(p_m(x_1{-}L)\Bigr)\,\sin\Bigl(p_m(x_2{-}L)\Bigr)\,
e^{-(x_1-x_2)^2/\alpha_c^2}\, ,
\label{BCS1}
\eeq
where $I_0(z)$ is the modified Bessel function.

Integrating then over $x_1$ and $x_2$ and introducing the
function
$$
g_c(p,k)=\frac{i\sqrt{\pi}\alpha_c}{2L^2(p{+}k)}
\Biggl\{
e^{-\alpha_c^2 k^2/4}
\biggl[ \biggl( e^{2i(p+k)L}+1 \biggr)
    {\rm erf}\biggl({ik\alpha_c\over 2}\biggr)
  - {\rm erf}\biggl({ik\alpha_c\over 2}{-}{2L\over \alpha_c}\biggr)
$$
$$
  - e^{2i(p+k)L}
    {\rm erf}\biggl({ik\alpha_c\over 2}{+}{2L\over \alpha_c}\biggr)
 \biggr]
+ e^{-\alpha_c^2 p^2/4}
\biggl[ \biggl( e^{2i(p+k)L}+1 \biggr)
    {\rm erf}\biggl({ip\alpha_c\over 2}\biggr)
$$
\beq
  - {\rm erf}\biggl({ip\alpha_c\over 2}{-}{2L\over \alpha_c}\biggr)
       - e^{2i(p+k)L}
    {\rm erf}\biggl({ip\alpha_c\over 2}{+}{2L\over \alpha_c}\biggr)
 \biggr]
\Biggl\}   \, ,
\label{gc}
\eeq
we arrive at the gap equation in the following form
$$
\Delta(p_m,p_{\perp})=-{1\over 2}
\sum\limits_{c=1}^2 \alpha_c^2\,(W_c-B_c-H_c+M_c)\,
e^{-\alpha_c^2 p_{\perp}^2/4}
\int\limits_0^{\infty} k_{\perp}\, dk_{\perp}
e^{-\alpha_c^2 k^2_{\perp}/4}
I_0\left({\alpha_c p_{\perp} k_{\perp}\over 2}\right)
$$
\beq
\times\sum\limits_n\,\frac{\Delta(k_n,k_{\perp})}
{\sqrt{\xi^2(k_n,k_{\perp})+\Delta^2(k_n,k_{\perp})}}\,
G_c(p_m,k_n)  \, ,
\label{gapeq}
\eeq
where
$$
G_c(p,k)={1\over 8} \, {\rm{Re}}\,\Bigl\{
  g_c(p{+}k,p{+}k)  + g_c(p{-}k,p{-}k)
+ g_c(p{+}k,{-}p{-}k)
$$
\beq
+ g_c(p{-}k,{-}p{+}k)
 - 2g_c(p{+}k,{-}p{+}k) - 2g_c(p{+}k,p{-}k) \Bigr\}\, .
\label{ggc}
\eeq

To reveal spatial behavior of the nonlocal pairing gap operator
the Wigner transform of the gap is very useful. It reads
\beq
\Delta({\bf R},{\bf k})=\int d{\bf s}\,
\Delta({\bf R},{\bf s})\, e^{i{\bf k}{\bf s}}\, ,
\label{wigner}
\eeq
where ${\bf R}={(\bf r_1}{+}{\bf r_2})/2$ and
${\bf s}={\bf r_1}{-}{\bf r_2}$.
In the case of slab geometry, the Wigner transform (\ref{wigner})
of the gap operator depends only on $X$ which is the component of
${\bf R}$ perpendicular to the surface. Considering
$\Delta(X,k_x,k_{\perp})$
only for $X>0$, since it is an even  function of $X$, one can
 easily obtain the following  series:
$$
\Delta(X,k_x,k_{\perp})={1\over \pi L}\, \theta(L{-}X)
\sum\limits_{n=1}^{\infty} \Delta(k_n,k_{\perp})
\Biggl[
  2\cos\Bigl(k_n(X{-}L)\Bigr) \frac{\sin\Bigl(2k_x(X{-}L)\Bigr)}{k_x}
$$
\beq
 - \frac{\sin\Bigl(2(k_x{-}k_n)(X{-}L)\Bigr)}{k_x{-}k_n}
 - \frac{\sin\Bigl(2(k_x{+}k_n)(X{-}L)\Bigr)}{k_x{+}k_n}
\Biggr]  \, .
\label{sum}
\eeq

In the bulk, the gap $\Delta$ depends mainly on the total
momentum $k{=}\sqrt{k_x^2 {+} k_{\perp}^2}$. Approximately,
this is true also for the surface region. Within this approximation, we
can treat the gap at the Fermi surface $\Delta(X,k_{\Fs}(X))$
as the Wigner transform $\Delta(X,k_x,k_{\perp})$ taken at $k=k_{\Fs}(X)$,
where $k_{\Fs}=\Bigl({3\pi^2\rho(X)}/2\Bigr)^{1/3}$ is the local Fermi
momentum. We define the gap-shape function as the gap at
the Fermi surface normalized to one at the center of the slab:
\beq
\chi^{\Go}_{\Fs}(X)=\Delta\Bigl(X,k_{\Fs}(X)\Bigr)/
          \Delta\Bigl(0,k_{\Fs}(0)\Bigr) \, .
\label{shape_gogny}
\eeq

The function $\chi^{\Go}_{\Fs}(X)$ is drawn
in Fig.~1 (a and b)
for four different values of the half-width of the slab:
$L=4,6,8$ and 10\,fm. In Fig.~1\,a the gap-shape function
is calculated for the Fermi momentum $k_{\Fs}=1.4232$\,fm$^{-1}$
corresponding to the Fermi energy $\varepsilon_{\Fs}=42$\,MeV,
while in calculation of Fig.~1\,b the value of Fermi momentum,
$k_{\Fs}=1.4894$\,fm$^{-1}$, corresponds to
$\varepsilon_{\Fs}=46$\,MeV.

\begin{wrapfigure}{L}{7.5cm}
\mbox{\epsfig{file=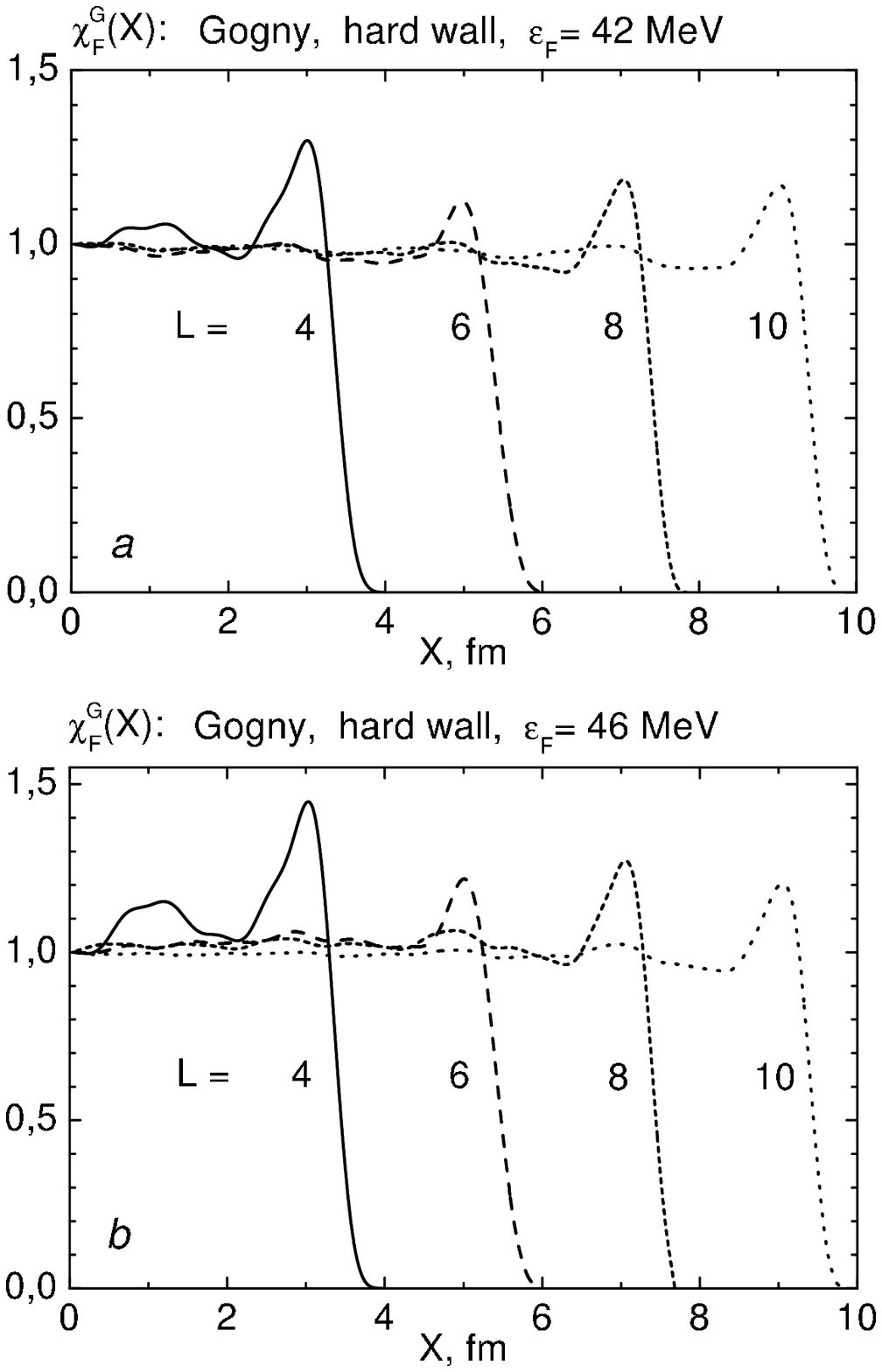,width=6.5cm}}
\caption{\small The
gap-shape function $\chi^{\Go}_{\Fs}(X)$ calculated in the model
of pairing with the Gogny force in the slab of nuclear matter
within the hard wall potential for $\varepsilon_{\Fs}=42$\,MeV
(panel a) and $\varepsilon_{\Fs}=46$\,MeV (panel b). The
half-width $L$ of the slab is given (in fm) by the numbers close
to the curves. }
\end{wrapfigure}

Let us discuss the salient
features of the  shape function
$\chi^{\Go}_{\Fs}(X)$. The main observation is that in the slab,
almost independent of its thickness,
the surface enhancement is not very different from
the semi-infinite matter case previously studied \cite{FS}.
The enhancement of the surface peak is rather moderate, not exceeding
20\% - 30\%. In fact it is not clear how much of
the surface peak comes from Friedel oscillations and how much
is a genuine enhancement of $\Delta$.
From this result and our previous study \cite{FS}
we are therefore inclined to conclude that at least with the Gogny
force pairing is only very moderately surface enhanced.
As to the dependence of the gap-shape function on the size of the slab,
in eq.~(\ref{shape_gogny}) there is a competition between two effects.
First, with diminishing of the size parameter $L$, the
peak value of the numerator $\Delta(X \simeq L)$ becomes higher.
Second, the denominator (the gap value in the center) also gets enhanced.
As a result, the $L$-dependence of the gap-shape function turns out
to be rather smooth.

Of course, finite nuclei are rather different from a slab
but qualitatively things should be similar. Some results for
finite nuclei with the Gogny force do exist \cite{Sn}
in the $Sn$-region but no definite conclusions
about the surface features of $\Delta$
can be drawn from this limited number of values.
In fact, in finite nuclei like the tin isotopes,
we should look at a great number of nuclei in the isotopic chain
because the behavior of $\Delta(r)$ can fluctuate a great deal
passing through the open neutron shell in question.

A detailed study of several long isotopes chains \cite{Fay}
with density dependent effective pairing forces confirmed
that, in the framework of a phenomenological approach,
it is rather difficult to distinguish between two opposite
possibilities, the volume pairing and the pairing with
pronounced surface enhancement. For this purpose, the analysis of
some binding energy characteristics, such as separation energies,
should be accompanied by the  study of the variations of nuclear radii
along the chain. The odd-even staggering phenomenon is especially
sensitive to the coordinate dependence of $\Delta$.
The microscopic calculation of $\Delta$ should help to solve
this problem.

Of course,
the conclusion may strongly depend on the employed pairing
force and below we will investigate the surface
behavior of pairing with a separable version of the
Paris force.

\vskip .5 cm
 \centerline {\bf 3. Pairing with the Paris force in a slab
of nuclear matter}
\centerline {\bf within a Saxon-Woods potential.}

\vskip .3 cm
Now we shall adopt the more realistic
one-dimensional Saxon-Woods potential $U(X)$
for a slab with the width of $2L$  symmetrical
around the origin $X=0$:
\beq
U(X) = \frac {U_0}{1 + \exp ((X{-}L)/d) + \exp (-(X{+}L)/d)}.
\label{sw}
\eeq

Here $U_0$ is the potential depth in the central region and
$d$ is the diffuseness parameter (to be more exact, the maximum
potential depth is $U(0) = {U_0} / \left(1 + 2 \exp (-L/d)\right)$).
The two parameters ($U_0{=}-50\;$MeV and $d{=}0.65\;$fm ) are taken to be close to
those of real atomic nuclei. The half-width parameter $L$ will be changed
to examine the size dependence of the effect under consideration.

To avoid a rather cumbersome resolution of the Bogolyubov equations
for the nonlocal gap \cite{BLSZ3}, as in \cite{BLSZ1}, we use
a powerful method \cite{KKC} (we refer to it as KKC) of solving
the gap equation for the case of a nonlocal interaction. This method
was originally suggested for infinite matter where
the gap $\Delta$ can be represented as a product
$\Delta(p) = \Delta_{\Fs} \chi(p)$ of the constant
$\Delta_{\Fs}=\Delta(p_{\Fs})$
and the "gap-shape" function $\chi(p)$ normalized to
$\chi(p_{\Fs})=1$.
Basically the KKC method is a transformation of the gap equation
to a set of two coupled equations: an integral equation for $\chi(p)$,
which is
almost independent of the value of $\Delta_{\Fs}$, and an algebraic
equation
for the value $\Delta_{\Fs}$. This significantly simplifies the solution of
the gap equation
in infinite matter. In \cite{BLSZ1} the extension of
the KKC method to nonzero temperatures \cite{KKC,K1} was used in
the case of semi-infinite nuclear matter where the spatial dependence
of the gap-shape function was also taken into account.
Within the KKC method for infinite matter, the temperature and momentum
parts of the gap function can be factorized as follows:
$\Delta ({\bf p},T)=\Delta_{\Fs}(T) \,\chi({\bf p})$.
In \cite{BLSZ1} it was supposed that a similar separation of the
temperature factor can be made for the semi-infinite system:
\beq
\Delta (x_1,x_2,k^2_{\perp};T)= \Delta_{\Fs}(T) \,\chi(x_1,x_2;k^2_{\perp}).
\label{ansatz}
\eeq
An additional advantage from using
the KKC method in this case comes from the possibility of finding
the normalization factor $\Delta_{\Fs}(T) $ by solving the gap
equation in infinite matter.

In this paper we use the same ansatz (\ref{ansatz}) for the
slab geometry. Unfortunately, in this case no direct
relation to infinite nuclear
matter exists and there is no evident way
to find the normalization factor without solving the Bogolyubov
equations. However, since the gap-shape function is of main importance
for an analysis of the surface enhancement of the gap, we do
not calculate here the normalization factor postponing the solution of
the Bogolyubov equation to a forthcoming  publication.

As far as in the case of the slab geometry all the equations
are very similar to those for semi-infinite matter, we write down
explicitly
only those which are necessary for explaining our calculations
and refer the reader to \cite{BLSZ1,BLSZ2} for details.
In symbolic notation, the gap equation has the form
\cite{AB,Schuck}:
\beq
\Delta(T)=\Veff\,A_0^s (T)\,\Delta(T)
\label{gap_ren}
\eeq
where $\Veff$ is the effective pairing interaction
acting in the model space $S_0$ in which the superfluid two-particle
propagator $A_0^s $ is defined.

The separable $3\times 3$ form \cite{Par1,Par2} of the
Paris potential \cite{Paris} is used
\beq
V({\bf k},{\bf k'}) = \sum\limits_{ij}\lambda_{ij} \,
g_i(k^2)\, g_j(k'^2),
\label{paris}
\eeq
where ${\bf k}$ and ${\bf k'}$ are the relative momenta
before and after scattering.
The effective interaction has a similar separable form
which, in notation of \cite{BLSZ2}, is as follows:
\beq
{\Veff}(x_1,x_2,x_3,x_4,k^2_{\perp},k^{\prime 2}_{\perp};E) =
\sum\limits_{ij}\Lambda_{ij}(X,X';E)\,
g_i(k^2_{\perp},x) \,g_j(k^{\prime 2}_{\perp},x'),
\label{veff}
\eeq
where $E{=}2\mu$, $X{=}(x_1{+}x_2)/2$, $x{=}x_1{-}x_2$, etc.,
and $g_i(k^2_{\perp},x)$ stands for the inverse Fourier transform
in the $x$-direction of the form factor $g_i(k^2_{\perp}+ k^2_x)$.
The gap-shape factor can be also written as
\beq
\chi(x_1,x_2;k^2_{\perp}) =
\sum\limits_i\chi_i(X)\,g_i(k^2_{\perp},x).
\label{chi}
\eeq

After substituting Eqs.~(\ref{ansatz}), (\ref{paris})~--~(\ref{chi})
into Eq.~(\ref{gap_ren}) at $T=T_c$ we obtain the following equation for the
components $\chi_i$:
\beq
\chi_i(X)=\sum\limits_{lm}\int \int dX_1dX_2\,
\Lambda_{il}(X,X_1;E)\,B_{lm}(X_1,X_2,E;T_c)\,
\chi_m(X_2),
\label{chi_i}
\eeq
where
\begin{eqnarray}
\lefteqn{
B_{lm}(X,X',E;T)=-\sum\limits_{n_1n_2}
\int\frac{d{\bf k}_{\perp}}{(2\pi)^2}\,\frac
{1{-}N_{\lambda_1} (T){-}N_{\lambda_2} (T)}
{E -\varepsilon_{\lambda_1} -\varepsilon_{\lambda_2} }
\times
}\qquad\qquad\qquad \qquad\qquad\qquad
\nonumber \\
& &{}\times G^l_{n_1n_2}(k^2_{\perp},X)\,
G^m_{n_1n_2}(k^2_{\perp},X'),
\qquad\qquad
\label{blm}
\end{eqnarray}
\beq
G^l_{n_1n_2}(k^2_{\perp},X)=\int y_{n_1}(X{+}{x\over 2})\,
g_l(k^2_{\perp},x)\,y_{n_2}(X{-}{x\over 2})\,dx,
\label{gl}
\eeq
\beq
N_{\lambda}(T)=\left({1{+}\exp{\frac{\varepsilon_{\lambda}{-}\mu}{T}}}
\right)^{-1},
\label{occup}
\eeq
In Eqs.~(\ref{blm})\,--\,(\ref{occup}),
$\lambda = (n,{\bf k}_{\perp})$,
$\varepsilon_{\lambda} = \varepsilon_n{+}k^2_{\perp}/2m$,
$\varepsilon_n$ and $y_n(x)$ stand for the energies and wave functions,
respectively, of the 1-dimensional Schr\"odinger equation
with the potential (\ref{sw}).

Following the recipe of Ref.~\cite{BLSZ2} for
the effective interaction $\Veff$
the propagator $A_0^s$ embodies all the single-particle states with
negative energies only. Thus, the summation over $n_1,n_2$ and the
integration over ${\bf k}_{\perp}$ are limited by the conditions:
$\varepsilon_{\lambda} <0, \; \varepsilon_{\lambda'} <0$.

The coefficients $\Lambda_{ij}$ obey the set of
integral equations
\begin{eqnarray}
\lefteqn{
\Lambda_{ij} (X_{12},X_{34};E) =\lambda_{ij} \delta(X_{12}{-}X_{34}) +
}
\nonumber \\
& &{} + \sum_{lm}\lambda_{il}\int dX_{56}\,B_{lm}(X_{12},X_{56};E)
\,\Lambda_{mj}(X_{56},X_{34};E),
\label{lambda}
\end{eqnarray}
where $B_{lm}$ are defined by an expression similar to
Eq.~(\ref{blm}),
but without the temperature factor, including the
states $\lambda_1, \lambda_2$
from the complementary subspace.

To simplify the calculations we used the local potential
approximation (LPA) which has turned out to be accurate
for semi-infinite nuclear matter \cite{BLSZ2}
and for nuclear slabs \cite{BLSZ4}.
The LPA prescription consists in using for the 2-particle
propagator $B$ of the complementary space
the local momentum approximation, very similar to the
Local Density Approximation (LDA) \cite{LDA1}, for each particle
separately:
$\varepsilon_n \to p_x^2/2m + V(X) $,
$\varepsilon_{n'} \to (p'_x)^2/2m + V(X) $.
This type of approach, where the individual particles are
treated in semiclassical approximation was used in
\cite{GD} for examining the response function.
This approximation has been shown to be
very accurate, if one is not interested in fine details.
It should be stressed that the LPA
is only applied to the equation for the effective
interaction $\Veff$, while no local approximation is
used in the renormalized gap equation (\ref{gap_ren}).
That is, the local approximation is used only for
two-particle states belonging to the complementary
space, for which the corresponding energy denominators in
eq.(\ref{blm}) are large. Therefore the individual contribution
of each state is negligible, and only the sum of a number of
such contributions is important. For such
a sum the semiclassical and local approximations
are expected to be accurate.
In this respect LPA is
different from the standard LDA, since in the latter
the local approximation
is used for all two-particle states.

Within the LPA, the exact values of $B_{lm}(X_1,X_2;E) $ are
replaced by the set of \linebreak $B_{lm}^{\rm\scriptsize
inf}(t,E;U[X])$ calculated for infinite nuclear matter put into
the homogeneous potential well of the depth $U[X]$. Here
$t{=}X_1{-}X_2$, $X{=}(X_1{+}X_2)/2$ are the relative and average
values of the CM coordinates. In the first step of the LPA
procedure we calculate the set of vectors $B_{lm}^{\rm\scriptsize
inf}(t,E=2\mu;U_i)\; (V_i{=}\delta V {\cdot} (i-1))$ for a fixed
value of the chemical potential $\mu$. In  the second step, for
each value of $(X,t)$, we find $B_{lm}^{\rm\scriptsize
LPA}(X_1,X_2)$ interpolating the values of $B_{lm}^{\rm\scriptsize
inf}(t;V_i)$ by the values of $V_i$ nearest to $V(X)$. After
substitution of $B_{lm}^{\rm\scriptsize LPA}(X_1,X_2)$ into
Eq.~(\ref{lambda}) we find the LPA prescription
$\Lambda_{lm}^{\rm\scriptsize LPA}(X_1,X_2)$ for the effective
interaction which should be substituted into the homogeneous
Eq.(\ref{chi_i}) for the gap-shape function. If the critical
temperature $T_c$ was known, this equation could be solved
directly. To find $T_c$ a more general integral equation must be
considered \beq \chi_i(X)= \lambda(T) \sum\limits_{lm}\int dX_1\,
K_{il}(X,X_1)\,\chi_m(X_1), \label{chi_lin} \eeq which involves
the eigenvalue $\lambda(T)$. Here the abbreviation
$K=\Lambda\,B^0$ is introduced for the kernel. The critical
temperature can be found from the evident condition
$\lambda(T_c)=1$.

The entire calculation scheme is similar to that for semi-infinite
matter except for some details. First, in the slab case we are dealing
with the discrete spectrum $\varepsilon_n$ in Eq.~(\ref{blm}).
Second, due to the obvious reflection symmetry of the slab
system in the $x$-direction, all the integral equations
under consideration can be readily reduced to a form including
positive $X$ only. Just as in \cite{BLSZ1}, instead of the
direct solution of Eq.~(\ref{chi_lin})
in the coordinate space, we use the Fourier expansion
within the symmetrical interval $(- L_0 ,L_0 )$,
\beq
\chi_i(X) = \sum\limits_n \, \chi_i^n f_n(X)\, ,
\label{fourier}
\eeq
where only the even functions must be retained, $f_n(x){=}
\cos(\pi n(X{-}X_c)/L_0)$.
The kernels $K_{ij}(X,X')$ of Eq.~(\ref{chi_lin}) are also
expanded in a double Fourier series. Finally we arrive at
a set of
homogeneous linear equations for the coefficients $\chi_i^n$ :
\beq
\chi_i^n= \sum\limits_{j=1}^3 \sum\limits_{n'=1}^N
\, K_{ij}^{nn'}\, \chi^{n'}_j,
\label{lineq}
\eeq
which can be solved by standard numerical methods.
Then the components of the gap-shape function are found
from Eq.~(\ref{fourier}).

\begin{wrapfigure}{L}{7.5cm}
\mbox{\epsfig{file=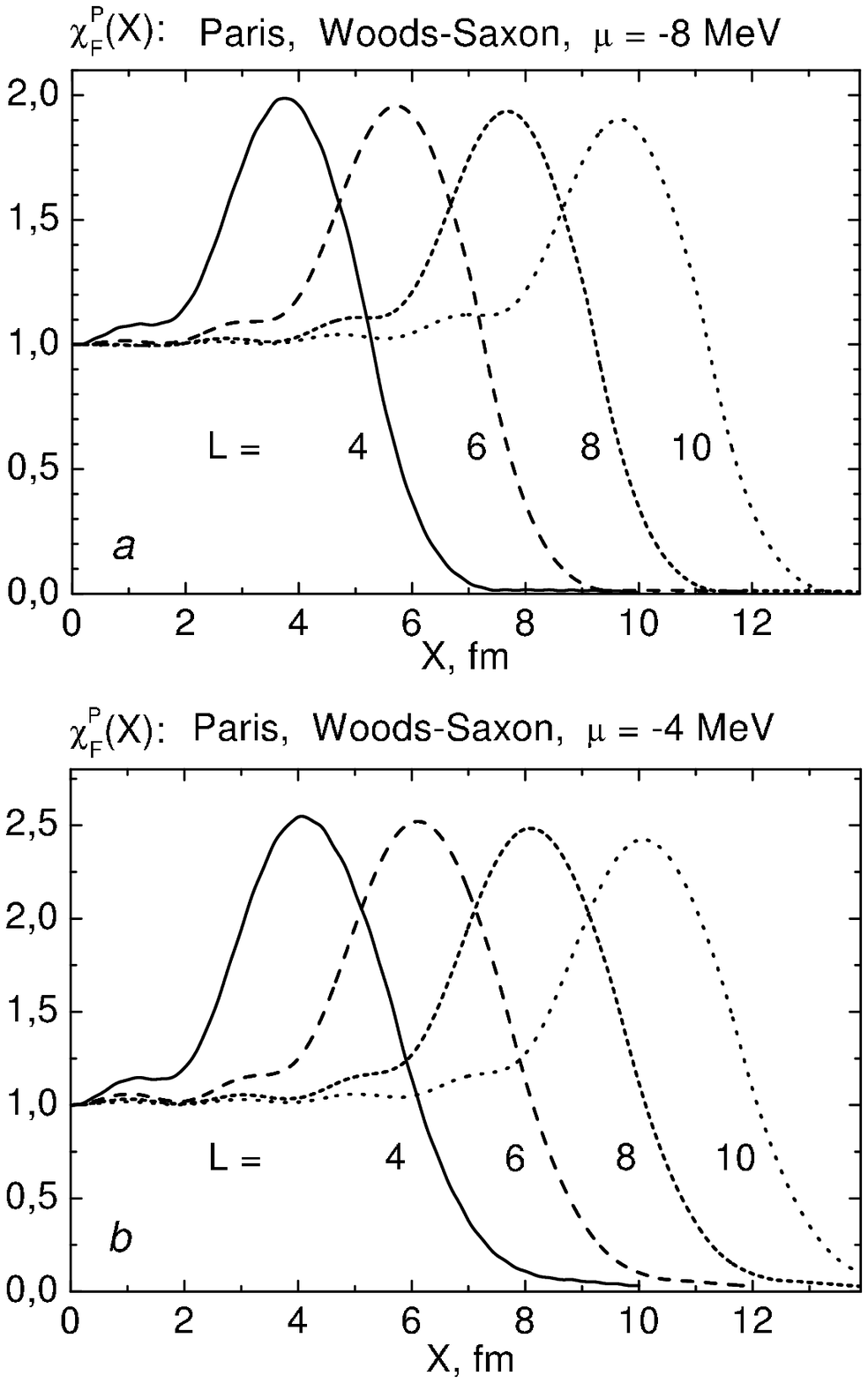,width=6.5cm}}
\caption{\small The gap-shape function $\chi^{\Pa}_{\Fs}(X)$
calculated in the model of pairing with the Paris force in the
slab of nuclear matter within the Saxon-Woods potential for
$\mu=-8$\,MeV (panel a) and $\mu=-4$\,MeV (panel b).
The half-width parameter $L$ is given in the same way
as in Fig.~1.}
\end{wrapfigure}

Instead of analyzing the separate components $\chi_i(X)$
for the separable form (\ref{paris}) of Paris force
it is more useful to display the local form of the gap-shape function
which enters the matrix elements of the gap for states nearby
the Fermi surface:
\beq
\chi^{\Pa}_{\Fs}(X) =  \sum\limits_i \, \chi_i(X) \,
g_i\Bigl(k^2{=}k_{\Fs}^2(X)\Bigr),
\label{chif}
\eeq
where $k_{\Fs}(X)=\sqrt{2m(\mu{-}U(X))} $ is the local Fermi
momentum ($k_{\Fs}(X)=0$ for \linebreak
$\mu{-}U(X) <0 $).
We calculated the gap-shape function
$\chi^{\Pa}_{\Fs}(X)$ for the same values of the half-width
of the slab, $L=4,6,8$ and 10 fm, as in the case of the hard wall
potential and for two values of the chemical potential
$\mu=-8$ MeV and $-4$ MeV. As the depth of the Saxon-Woods
potential was taken $U_0=-50$\,MeV, these two values of $\mu$
correspond just to the same values of the Fermi energy as in
the hard-wall case. The results are shown in Fig.~2 a and b.

One observes a surface bump which is much more pronounced than in the
previous model with the Gogny force and the hard wall potential.
The enhancement is now around 80\% - 100\% and, as for the Gogny
force, it is quite similar to the semi-infinite matter case \cite{BLSZ1}.
Two values of the chemical potential
$\mu=-8$\,MeV and $\mu=-4$\,MeV
have been chosen which in account of the
depth of the Saxon-Woods potential of $V_0=-50$\,MeV correspond
precisely to the Fermi energies
$\varepsilon_{\Fs}=42$\,MeV and $\varepsilon_{\Fs}=46$\,MeV,
respectively, of the previous model.
Going from $\mu=-8$\,MeV to $\mu=-4$\,MeV
a rather important increase of the enhancement of the order
of 30\% is observed which is much larger  than in
the case of the hard-wall potential with the Gogny force.
On the other hand, there is little variation with the
thickness of the slab, the maximum of $\Delta$ being a couple
of percent larger for small slab size. This situation is analogous
to the previous model.

The surface effect in $\Delta$ for the Paris force can be
qualitatively explained from the properties  of the effective interaction
$\Veff$
which were analyzed in \cite{BLSZ5}. There it was  found
that $\Veff$ undergoes a sharp variation in the surface region, from
almost zero in the bulk to very strong attraction
in vacuum. In the asymptotic region,
the latter coincides with the off-shell $T$-matrix
of free $NN$-scattering  $T(E=2\mu)$
which exhibits a resonant behavior at small $E$. The
strong surface attraction and the  sharp variation in the surface region
are mostly responsible for the surface effect of the gap.
The $\mu$-dependence of the surface effect can be explained
by the increase of the jump
$\delta {\Veff}$ from inside to outside as
$|\mu|$ is decreasing.
There are two reasons for such an increase.
The first one is the
$k^2$-dependence of the form factors in Eqs.(\ref{paris}), (\ref{veff})
leading to a reduction of $\Veff$
in the inner region with increasing values of $k_F$.
The second one is a pole-like behavior of $T(E)$ at small $E$
which results in an increase of $\Veff$
with decreasing $|\mu|$ in the exterior, due to the
approach to the virtual pole.
One sees that both effects work in the
same direction resulting in strengthening the surface effect
at small values of $|\mu|$.

\vskip .5 cm
\centerline {\bf 4. Discussion, and conclusions.}
\vskip .3 cm

In this work we continued our effort to understand the surface
behavior of the nuclear gap. Previous investigations
considered semi-infinite nuclear matter
embedded in i) a hard wall potential with the Gogny force \cite{FS}
and ii) a potential of Saxon-Woods shape with the separable
version of the Paris force \cite{BLSZ1}. In both cases a surface
enhancement was found but which is relatively modest in view of what one
could expect from LDA.
In the present study we addressed the question whether finite size
effects can strongly alter this situation and repeated the
former calculation \cite{BLSZ1,FS} in a slab configuration.

In the first case of pairing with the Gogny force (D1) within
the slab of nuclear matter, the hard wall potential allows
to perform most part of calculations analytically. The
second rather sophisticated case demanded a lot of numerical
effort due to the use of the Paris interaction and the
Saxon-Woods shape of the mean field potential. In both
cases a noticeable surface effect for the pairing gap was
obtained of the same order of magnitude as it was previously found in
semi-infinite nuclear matter.
The shape of  the gap in coordinate space
turned out to be qualitatively similar
in both cases, with a significant surface enhancement.
For the value of the chemical potential $\mu{=}-8\;$MeV which
simulates  stable atomic nuclei, the enhancement is of
the order of 30\% for the first model and is almost 100\%
for the second one.
A general feature of both models is the rather smooth dependence
of the enhancement
on  the slab thickness which is approximately 10\% in the first case
and only 5\% in the second one. In both cases, a $\mu$-dependence
of the surface effect is found: the
enhancement coefficient increases as the absolute value of the chemical
potential $|\mu|$ decreases. The latter effect is more pronounced
for the Paris interaction and Saxon-Woods potential reaching
30\% with diminishing  $|\mu|$ from 8 MeV to 4 MeV.
For the Gogny force and box potential, the corresponding $\mu$-effect
is approximately 10\%.

To understand in detail the possible common physical origin of
this  surface enhancement as well as the mentioned differences,
it is instructive to consider the gap equation in the form
(\ref{gap_ren}) in which $\Delta$ is expressed in terms
of the effective interaction $\Veff$ given by Eq.~(\ref{veff}).
Properties of the effective interaction generated by the Paris force
were analyzed in \cite{BLSZ5}, where it was found to undergo
a sharp variation in the surface region.
It is this sharp variation
that is mostly responsible for the surface effect in the gap.
It is natural that in the case of the hard wall box potential
the influence of the surface interaction is smaller which makes
the surface enhancement weaker.

The $\mu$-dependence of the surface effect is explained
by the increase of the jump
$\delta  {\Veff} =
{\cal V}_{\rm\scriptsize eff}^{\rm\scriptsize in} -
{\cal V}_{\rm\scriptsize eff}^{\rm\scriptsize ex}$
with decreasing $|\mu|$.
This increase is caused by two reasons. The first one is
the strong $k^2$-dependence of the form factors in Eq.(\ref{paris})
leading to a reduction of
${\cal V}_{\rm\scriptsize eff}^{\rm\scriptsize in}$
for the larger values of $k_{\Fs}$ in the bulk.
The second one is an increase of
${\cal V}_{\rm\scriptsize eff}^{\rm\scriptsize ex}$
with decreasing $|\mu|$ caused by
the pole-like behavior of the $T$-matrix at
small $E$. Thus, both reasons jointly work
towards making the surface effect stronger at small
values of $|\mu|$. Qualitatively, the two reasons should
work also for Gogny force, but the hard wall potential
strongly suppresses the second reason, diminishing  the surface
effect itself and its $\mu$-dependence as well.
Finally, it should be mentioned that
a large value of the coherence length of pairing which is
comparable with the size of the slab makes all the effects
under considerations rather smooth. In particular, it results in
the weak dependence of the surface enhancement of the pairing gap
on the slab thickness.

It is worth to point out that the mechanism of the
surface enhancement in $\Delta$ considered in this paper
is different from the one of ref. \cite{bort}.
The latter is related to a contribution to the
effective pairing interaction of the virtual exchange by
collective surface vibrations. In a consistent description
of the surface behavior of the pairing gap in nuclei these two
effects should be considered on equal footing.

\vskip 0.5 cm

This research was partially supported by Grants
No.~00-15-96590 and No.~00-02-17319 from the Russian
Foundation for Basic Research.
Two of the authors (E.E.S. and M.V.Z.) thank
INFN (Sezione di Catania and LNS) and Catania University for hospitality
during their stay in Catania.

\end{document}